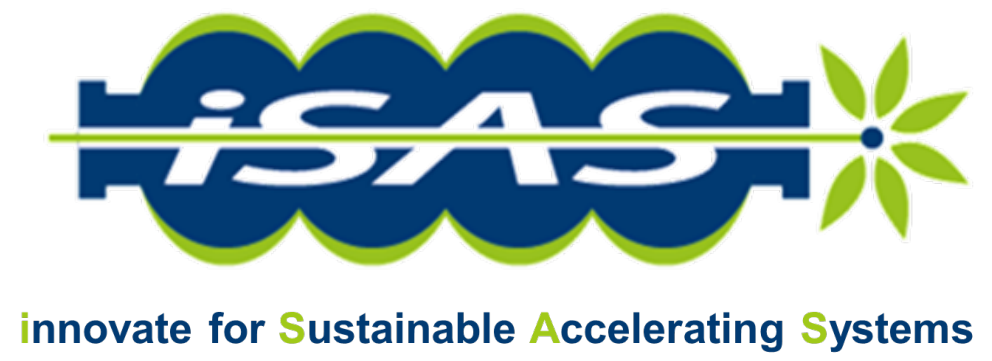

# Energy-saving metric of the iSAS project

This document presents the energy-saving metric of the project *Innovate for Sustainable Accelerating Systems* (iSAS), funded by the EU under its program HORIZON-INFRA-2023-TECH-01 via grant agreement n°101131435 (milestone 9.5).

**Author list:**

**M. Baylac (CNRS/LPSC), F. Bouly (CNRS/LPSC), J. Branlard (DESY), K. Canderan (CERN), J. D'Hondt (Nikhef), P. Duschene (CNRS/IJCLab), Y. Gómez Martínez (CNRS/LPSC), J. Knobloch (HZB-univ. Siegen), A. Neumann (HZB), V. Parma (CERN), C. Pira (INFN-LNL), H. Saugnac (CNRS/IJCLab), C. Schmidt (DESY), A. Stocchi (CNRS/IJCLab)**

## *Table of contents*

Funded by the European Union

Funded by HORIZON-INFRA-2023-TECH-01 via grant agreement n°101131435

# 1. INTRODUCTION

Several technologies are developed within the iSAS program towards sustainable particle accelerators [1]. Task 9.5 is dedicated to defining how to quantify the energy savings of these technologies. To meet the associated milestone, the present report aims to provide a verifiable metric for energy saving performances of the different technologies. This metric will be filled in at the end of the project, with the equipment available and operational at that time.

For this energy saving metric, the selected approach is focused on **operational costs**. We will evaluate and compare the power consumption with and without the « iSAS technology » under similar conditions:

- We will measure (or calculate) the electrical power consumption
- We will compare this consumption with and without the iSAS technology (or iSAS option)
- The comparison will be performed for different use cases (specific to each technology) but always under similar conditions for each technology.

The metric to be defined for all the iSAS technologies, that is:

- FE-FRT (Ferro-Electric Fast Reactive Tuners)
- LLRF (Low Level Radio Frequency) control system
- $Nb_3Sn$ on Cu films cavity
- Couplers
    - FPC (Fundamental Power Coupler)
    - HOM (Higher Order Mode)
    - BLA (Beam Line Absorber)

**The proposed metric has been submitted to and approved by the advisory board of iSAS (April 2025).**

In addition, a consumption model of the PERLE ERL will be developed to evaluate the accelerator efficiency.

# 2. FE-FRT

WP1 aims to develop novel a fast-tuning system compensating detuning by mechanical vibrations or beam induced transients, which both are usually compensated by adding a significant power overhead. Ferro-Electric Fast Reactive Tuners (FE-FRT) tuners are an alternative to classic mechanical cavity tuner [2]. They form a coupled system of cavity and a usually coaxial line with ferro-electric material. The permittivity of this material can be changed by applying HV, thus generating a change of impedance allowing to control the resonance frequency of the coupled system on a very fast time scale (~100 ns). FE-FRTs require an additional RF port to install on the cavity, but circumvent the complexity of control algorithms needed to damp the highly resonant mechanical-RF cavity-tuner system. FE-FRTs should of course not compromise the stability and reliability of operation, but could even better improve it beyond current state of the art.

FE-FRT will be developed to be tested with a 2-cell 1.3 GHz cavity and a fully equipped nine cell TESLA/XFEL cavity, including a tunable high-power coupler and classic mechanical tuning system. The metric to estimate the savings is the invested RF power to drive the cavity at a given field (peak and average power) for microphonics compensation, which will be converted to wall-plug power.

Funded by the European Union

Funded by HORIZON-INFRA-2023-TECH-01 via grant agreement n°101131435

The study case is a TESLA cavity (1.3 GHz) in CW mode at 16-20 MV/m in 3 configurations:

- Reference test 1: with a piezo control at a conservative loaded quality factor (e.g. 1e7)
- Reference test 2: with a piezo control at a higher loaded Q of (e.g. 5e7)
- FE-FRT demonstration: with FE-FRT, demonstrate micro-phonics detuning compensation and reachable, field stable, and aim for highest loaded Q. Measure the RF power level and dissipated power in FRT.

**The energy saving metric for the FE-FRT will be determined by comparing the RF power invest with the FE-FRT and without the FE-FRT (with the piezo control) at identical values of loaded Q**.
Expected savings in RF power should be significant. The required power to operate a cavity at voltage scales with the square of the cavity bandwidth in low beam loading machines (microphonics case). For example, fig. 1 illustrates that a factor of 10 reduction in RF power is possible when reducing the cavity detuning from 40 Hz down to 5 Hz, which is reachable with a FE-FRT.

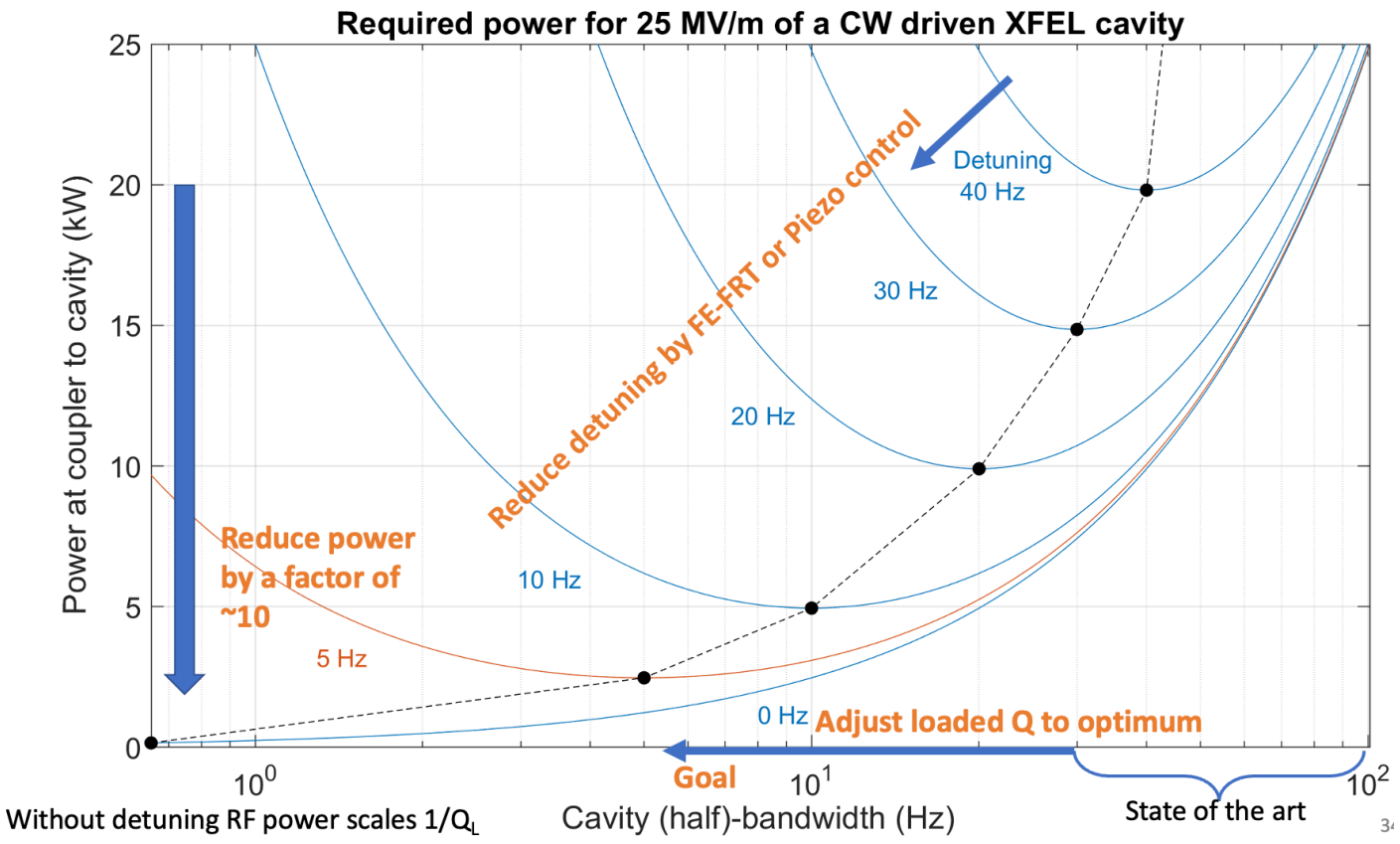

*Figure 1 : Required RF power to reach 25 MV/m with a CW driven XFEL cavity versus half-bandwidth (from iSAS proposal)*

State of the art of piezo control can reach a loaded Q of $Q_L$=5e7 (bandwidth 13 Hz) and a residual microphonics of 10 Hz peak: this requires about 3 kW forward power for a gradient of 20 MV/m, foreseen on a test stand. Comparatively, FE-FRTs can operate at $Q_L$=2e8 (bandwidth 3.25 Hz) and reduce peak detuning to 1 Hz, thus requiring only 530 W forward power. On the other hand, for machines operating at lower Q and larger bandwidth, as of higher peak detuning, the gain factor in bandwidth, and thus in power saving, is enlarged. One can estimate that by replacing the piezo control by FE-FRT, the expected reduction in cavity bandwidth, while maintaining a stable and operable system, is estimated around 4-20 such that the expected reduction in RF power for microphonics compensation is expected around 5-40.

## 3. SMART LLRF

WP2 aims to develop an optimized Low-Level Radio Frequency (LLRF) system, in order to reduce the RF power required to drive the cavities, thanks to an efficient control of cavity field and detuning. To minimize the RF power, cavities must operate at a quality factor $Q_{ext}$ as high as possible, at a narrower bandwidth. But these conditions make resonance control extremely challenging and will only reduce the RF power needs if the resonance can be properly guaranteed. So, the challenge is to find the highest $Q_{ext}$ ($\sim Q_L$) while meeting resonance control goals and without compromising operability or reliability. AI will be integrated in this smart LLRF.

Since a LLRF system is constantly required for cavity field control, the system cannot be switched OFF for comparison purposes. Yet, **options of the optimized LLRF** can be switched ON/OFF or adjusted to evaluate their impact on power consumption. Several factors of the LLRF tuning system impact the consumption of the accelerator (value of coupling, shape of the modulator pulse, efficiency of the frequency control, type of amplifier, drain voltage ...) so several metrics/curves will be used (table 1).

*Table 1 : Proposed metric to estimate energy saving of the smart LLRF system*

| Metric | How to measure | Applies to task |
|---|---|---|
| **Instantaneous AC consumption** | Measure the accelerator AC consumption with and with the applied LLRF optimization for a direct comparison (at the modulators for example) | 2.2, 2.3, 2.4 |
| **RF power usage** | Compare the peak and the integrated forward RF power with and without the LLRF optimization | 2.2, 2.3, 2.4 |
| **AC-to-RF efficiency** | Compute the AC-to-RF efficiency of the high-power source (i.e. SSA) with and without working point optimization by the LLRF | 2.5 |
| **Accelerator up-time or trip recovery time** | Provide examples where the accelerator up-time or trip recovery time has been improved following the implementation of the LLRF supervisory control and fault diagnostics. | 2.5 |

Different measurements will be performed at different facilities (DESY, HZB and CNRS) offering different test options (CW, pulsed, presence of beam, narrow bandwidth cavities, etc.…) which measurements will benefit from to cover multiple operational cases.

Optimization using the LLRF system can lead to significant savings in the electrical consumption of an accelerator. For example, the step-wise optimization of the modulator pulse shape at the EU-XFEL led to a significant reduction in instantaneous accelerator power consumption by more than 1 MW (fig. 2). This improvement translates into cost savings of several million euros on the yearly electrical bill.

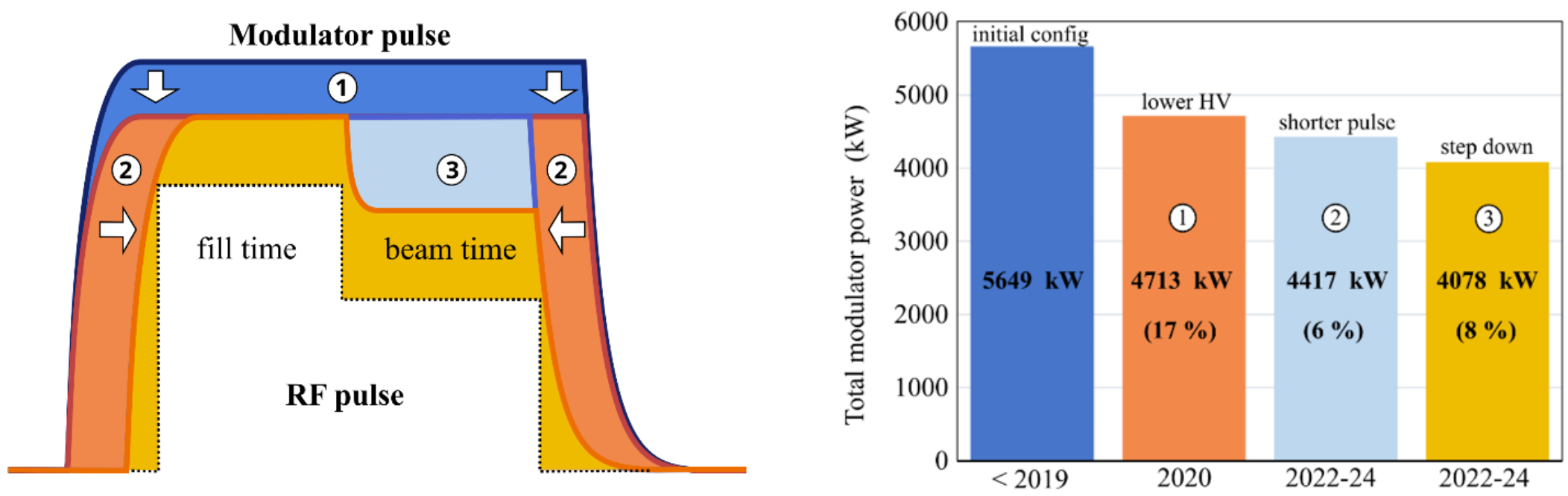

*Figure 2: Successive optimizations of the modulator pulse at EU-XFEL and corresponding power savings [3].*

A significant saving in RF power is expected for the smart LLRF system developed within iSAS. Simulations will be performed in order to evaluate the expected power saving of each of the four proposed metrics. For example, using the LLRF to optimize the working point of a GaN solid-state power amplifier can bring its efficiency up from 55% to 65%. Increasing the external quality factor ($Q_{ext}$) of a cavity operating at 20 MV/m from $Q_{ext}$ = 1e7 to $Q_{ext}$ = 4e7 can reduce the required power from 10 kW down to 4 kW, as illustrated in Fig. 3.

Funded by the European Union

Funded by HORIZON-INFRA-2023-TECH-01 via grant agreement n°101131435

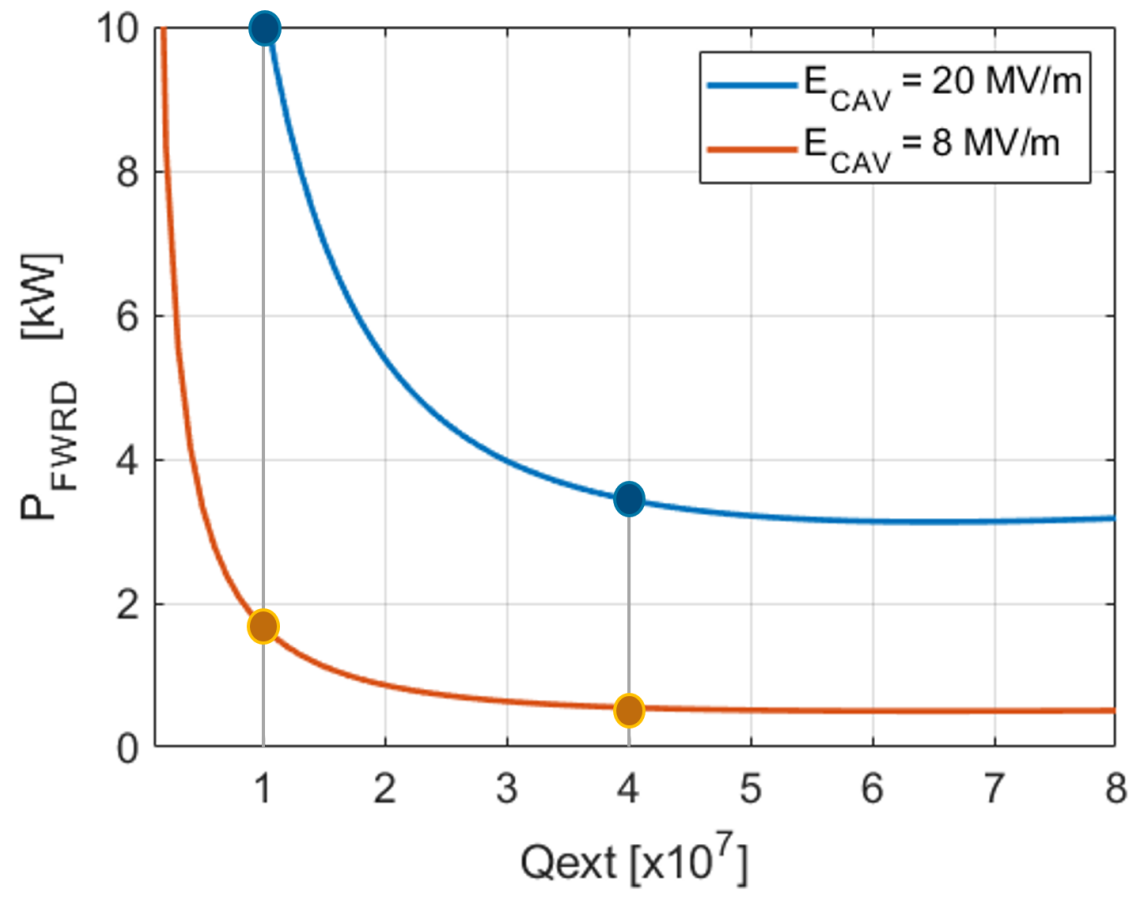

*Figure 3: Forward power required to reach a given accelerating gradient as a function of Qext.*

## 4. THIN FILMS ON CU CAVITY

WP3 explores coatings of $Nb_3Sn$ thin films on copper substrate in order to operate the RF cavities at temperature higher than 2K, thus reducing dramatically the required cryogenics power. To match the excellence performances obtained with cavities made of bulk Nb at operated 2K, thin film coatings must minimize flux trapping and ensure sufficient mechanical strength to allow for cavity tunability. WP3 will investigate these R&D topics on different types of samples and eventually with an operating cavity at 4.2 K.

A cavity in operation generates losses ($\propto 1/Q_0$) which must be dissipated by the cryogenic circuit. Given the coefficient of performances (COP) of the cryogenic system, dissipating these losses requires a large amount of power: about 800 W of power at room temperature is needed to account for 1 W of power dissipated at 2K. The efficiency, or quality factor, of the cavity is driven by its surface resistance $R_s$ via $Q_0 = G/R_s$ (with G, the cavity geometry factor). The surface resistance depends on several parameters (sputtering coating parameters, copper substrate surface treatments, interlayer, post-coating treatments). It comprises a temperature-dependent term (BCS) and the residual resistance: $R_s = R_{BCS}(T) + R_{res}$. The BCS term can be calculated as a function of temperature, therefore the residual resistance $R_{res}$ can be deduced from a usual measurement of $Q_0$ at a given temperature.
The energy required to operate a multi-cell cavity can be characterized by the power dissipation of the cavity cells at a given accelerating field: $P_{acc}$ per cell. This parameter takes into account the performance of both the material ($R_s$) and the cryogenics system (COP).

**To evaluate the energy saving of the thin film coatings, the proposed metric is the dissipated power: $P_{acc}$ per cell @ 1 MV/m.**
Determining this dissipated power for a $Nb_3Sn$ coated cavity requires a multiple-step methodology:

1. Measure $Q_0$ *versus* $E_{acc}$, via an RF test, to deduce the surface resistance $R_s$ at a given gradient
2. From this surface resistance evaluation of $Nb_3Sn$, estimate the power dissipation at 4.2K
3. Comparison with bulk Nb at 2K at same gradient by estimating the power dissipation at 2K.

**This evaluation will comprise 3 study cases:**

- Quadrupole resonators QPR (400 MHz, 800 MHz)
- 1.3 GHz cavity
- new cavity optimized at the end of ISAS.

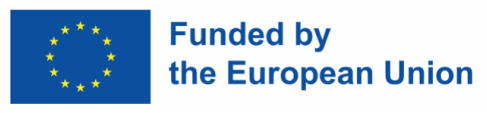

Funded by HORIZON-INFRA-2023-TECH-01 via grant agreement n°101131435

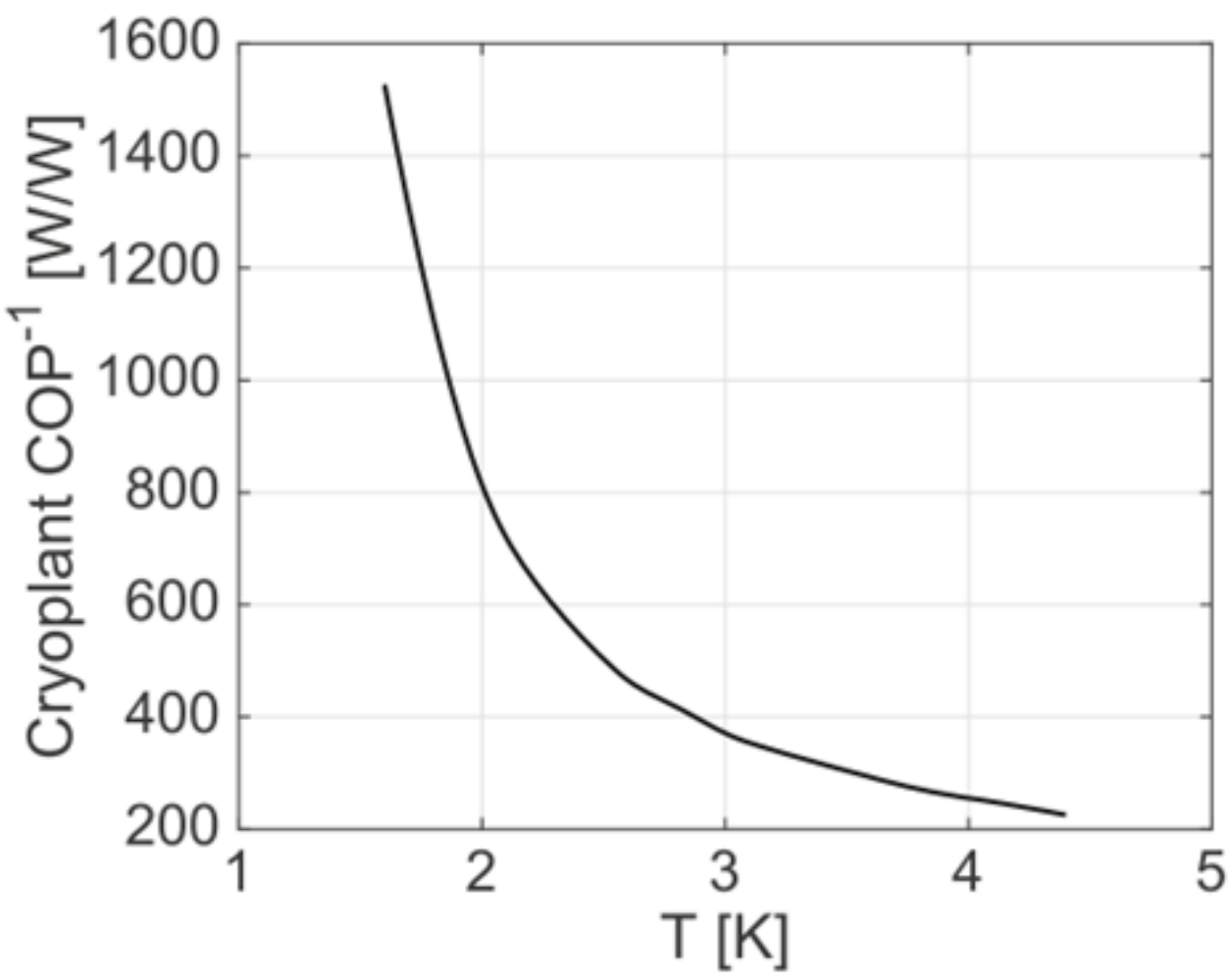

*Figure 4: Typical cryogenic plant efficiency given as inverse coefficient of performance (COP-1 indicates how many watts of wall power are required to remove one watt of heat) as a function of temperature. Data accounts for Carnot efficiency and deviation of a realistic plant from Carnot [4].*

With material having a higher critical temperature, it is envisaged that SRF cavities be cooled at 4.2K while maintaining high quality factor $Q_0$ and accelerating field $E_{acc}$. Considering the COP of the cryogenics system (≃240 at 4,2 K *versus* ≃800 at 2K, as seen in fig. 4), running at 4.2 K instead of 2K would reduce the power consumption by a factor of 3.

The expected dissipated power per cell ranges between 1 – 3.5 W per cell @ 1 MV/m ($R_{res}$ = 5-30 nΩ) as shown in figure 5. Thus, a $Nb_3Sn$ cavity operated at 4.2 K outperforms a Nb cavity at 1.8 K, even at residual resistance values up to 5-6 times higher.

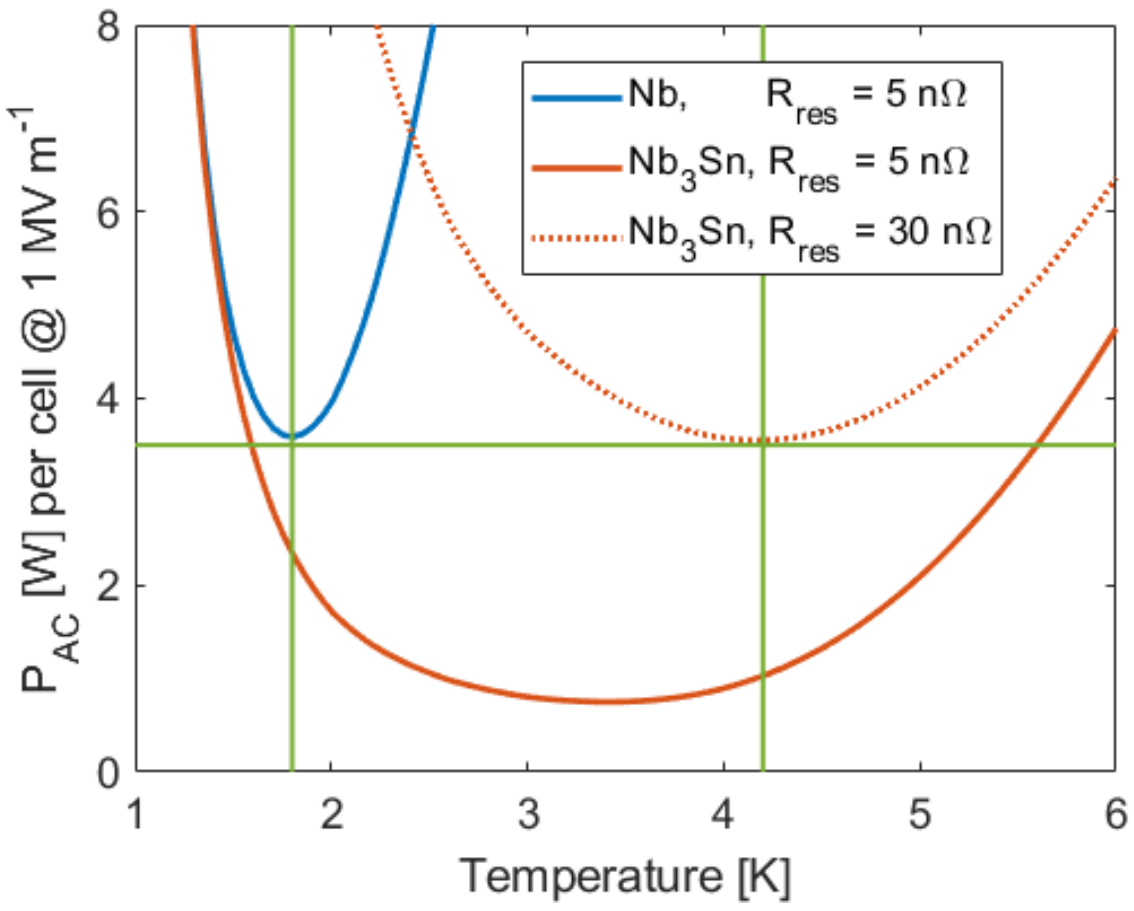

*Figure 5: Calculated AC power dissipation of Nb and Nb3Sn cavity cells versus temperature at 1.3 GHz [5]*

Funded by the European Union

Funded by HORIZON-INFRA-2023-TECH-01 via grant agreement n°101131435

## 5. COUPLERS

Regarding couplers, different technologies and devices are developed within WP4 at 800 MHz for the PERLE cryomodule:

- **FPC** (Fundamental Power Coupler): coupler to introduce the power to excite the fundamental mode of the cavity via an antenna, in order to accelerate the beam,
- **HOM** (Higher Order Mode): coupler used to damp the HOMs trapped in the cavity and deposit them into loads at higher temperatures. Even a moderate fraction of HOMs dissipated in the cryomodule's cold mass, will increase significantly the grid power for cryogenics. Moreover, improper extraction of HOMs can strongly limit the beam intensity due to beam instabilities.
- **BLA** (Beam Line Absorber): used to damp HOM propagating through the beam pipes by absorbing directly their power.

The cooling strategies for these components aims at reducing the heat loads to the cryogenic bath, while preserving the functional RF requirements.

Moreover, valuable feedback to further optimize the FPC system can be provided by analyzing the data from the ESS cryomodule (WP5).

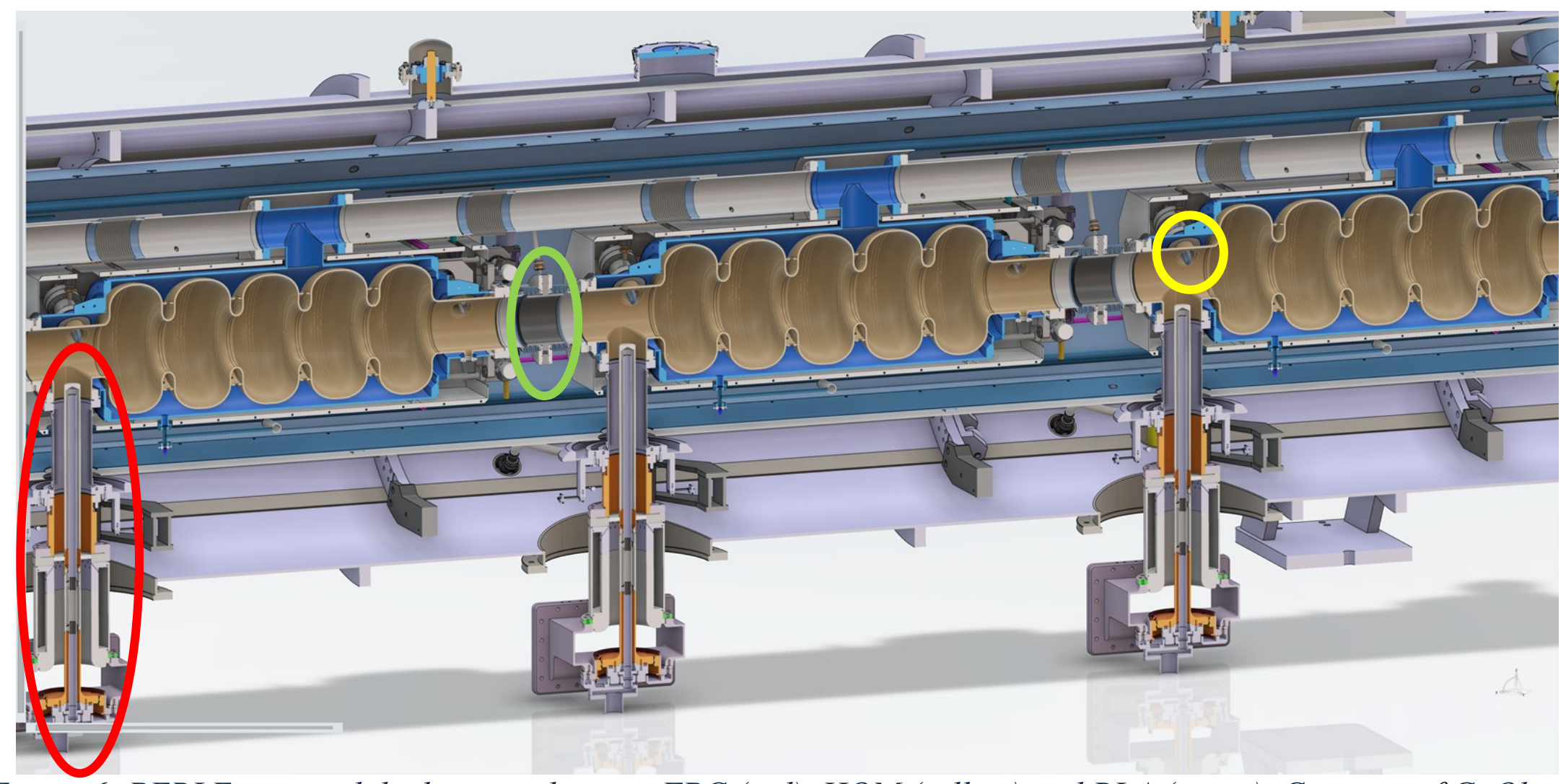

*Figure 6: PERLE cryomodule design indicating FPC (red), HOM (yellow) and BLA (green). Courtesy of G. Olivier (CNRS/IJCLab)*

**For the metric, we propose to calculate the electrical power consumption of the cryogenic system for different cases (defined below).** Optimizing the electric consumption for cryogenic process requires to minimize the exergie at 300K. The exergy at 300K is $exergy_{300K} = 300 \times (S_{out} - S_{in}) - (H_{out} - H_{in})$, where H and S correspond respectively to enthalpy and entropy, *in and out* being the input and output states of the processed fluid. The exergy depends mainly on the heat loads and the temperatures involved in the cooling process. It requires fine thermal simulations (finite elements 3D simulations) and choosing the optimized cooling processes. The process temperatures and pressures are imposed by the cryoplant (in our case 50K, 5K and 2K). We present here the electrical power optimization taking into account the Carnot efficiency only. One should note that the real overall efficiency of a cryoplant must be evaluated by using a Carnot efficiency, which depends on the processed fluid temperature, the cryoplant design and its overall cryogenic power. For instance, large state-of-the-art cryoplants have a Carnot efficiency of 30% of @ 50 K and 15% of @ 2 K, whereas for smaller cryoplants, like for

PERLE, the Carnot efficiency at 2K may be around 10%. In the following, we however use only the Carnot efficiency as it leads to an optimized choice for the cooling process, independently of the real cryoplant efficiency. The final estimation of the plug power will be finalized using the overall real efficiency of the actual cryoplant used for PERLE.

In addition, it must be taken into account that, when adding new components, the complexity of this extra component must also be considered: for example, the addition of a BLA in the cryomodule must be balanced against the risk of pollution induced by this extra material in the vicinity of cavities.

#### i. Fundamental power coupler

At 800 MHz, the FPC is based on the SPL coupler design [6], this design is already very efficient for large RF power. The outer conductor is the same used in the ESS cryomodule, a double-walled heat exchanger actively cooled with 23.1 mg/s of supercritical helium spilled from the supply line at 5K. With the current geometry and mass flowrate, the Reynolds number corresponds to a laminar regime. A turbulent regime was investigated to see if the heat transfer could be further improved with this design. Given the fixed geometry of the heat exchanger, a turbulent regime can be obtained only by increasing the mass flowrate. This option was proven inefficient by thermal calculations.

**For the metric, two cooling schemes will be studied and their cooling power consumption compared. For these two schemes, we consider, as a first approximation, that the solid conduction heat from the cold part of the FPC external conductor to the 2K bath is equal for the following both cases:**

**a) Two-loop coupler (heat exchangers at 50 K and 5 K)**

The RF power dissipation and solid conduction heat to the coupler's external conductor is evacuated with 2 isothermal loops. The design is easy and cheap to build, but the cooling efficiency is not optimal. The electrical power for cooling is estimated by:

- $P_{\text{elec Carnot}} \propto m'_{50K} \times 300 \times \text{exergy}_{50K\ loop} + m'_{5K} \times 300 \times \text{exergy}_{5K\ loop}$ (m': mass flow)

**b) One-loop coupler (heat exchanger from 5 K to 300 K)**

The RF power dissipation and solid conduction heat to the coupler's external conduction is evacuated through a heat exchanger with the fluid enthalpy from 5K and 300K. Energetically, it is much more efficient:

- $P_{\text{elec Carnot}} \propto m' \times 300 \times \text{exergy}_{5K \rightarrow 300K\ loop}$

With a given RF design, cooling scheme optimizations include materials to reduce the solid conduction (stainless steel body) and RF losses (copper coating). The one loop cooling scheme is selected as it is more energetically more efficient, about 2,5 times more than the two-loop scheme.

It is important to note that the energy saving obtainable with the one-loop cooling scheme with helium is strictly connected with the helium inlet temperature in the double walled tube, which should be as close as possible to the design value of 5K.

Moreover, studying the ESS data can help further optimization of the efficiency of the FPC. Based on the feedback from the ESS data, thermal calculations show that the major source of heat load to the cryogenic bath and power consumption related to the coupler is due the high temperature of the inlet cryoline to the double walled tube. Data from the ESS cryomodule indicate an inlet temperature to the FPC outer conductor of 9 K, for a design value of 5K, and as a consequence, increase the solid conduction heat transfer to the 2K bath. A major electrical power saving (in the kW range) is expected with a reduced conduction to the cryogenic bath by improving the thermal insultation on this inlet line feeding the FPC.

### ii. Higher order mode coupler

Several options of HOM couplers have been studied at 800 MHz [7]. The selected scheme for PERLE uses 2 Hook HOM couplers per cavity. **For the metric, two cooling schemes will be studied and their cooling power consumption compared:**

#### a) HOM coupler cooled at 2K (HOM@2K)

The RF load on the antenna is dissipated at 2K. It requires only one cooling tube for He II cooling channel (He II superfluid property). The electrical power for cooling is estimated by:

- $P_{elec\ Carnot} \propto 150 \times Q_{2K}$ ($Q_{2K}$: heat at 2K)

#### b) HOM coupler cooled at 5K (HOM@5K)

A higher RF load on the antenna is dissipated at 5K as a higher surface resistance of the material is expected. At this temperature, the heat transfer between the conductor and the He I is governed by convection and boiling rather than by conduction. Therefore, a circulation cooling loop with input and output tubes is required. An additional circuitry for the 5K fluid must be implemented inside the cryomodule, leading to additional cryogenic power:

- $P_{elec\ Carnot} \propto (60 \times Q_{5K} + 150 \times Q_{2K\ solid\ conduction})$ ($Q_{2K\ solid\ conduction}$: heat at 2K by solid conduction)

HOM optimizations include geometrical design: RF optimization (to reduce the RF losses) and thermal design, flange type and material, heat interception on the RF wire as well as materials to reduce the surface resistance and consequently the RF losses.
For High RRR Nb at 5K, the surface resistance is several orders of magnitude higher than at 2K. Thus, **a cooling at 5K (scheme b) could be energetically efficient only with superconducting materials other than Nb,** for instance $Nb_3Sn$ but it would require a R&D program.

It is important to mention that cooling at 5K can compromise the functionality of the HOMs. In case of unexpected loads, the antenna could reach a temperature exceeding the transition temperature and thus lose its superconductivity. Therefore, even if a cooling at 5K is more efficient, the 2K option may still be preferred thanks to a higher reliability against unexpected loads.

### iii. Beam Line Absorber

At 800 MHz, the BLA studies are underway to define the optimum BLA to installed in the PERLE cryomodule based on existing design [8]. **For the metric, two cooling schemes will be studied and their cooling power consumption compared:**

#### a) Without BLA inside the cryomodule (BLA@300K):

The power dissipated by the HOMs propagating through the beam pipes inside the cryomodule (CM) is evacuated at 2K. It is very difficult to simulate and estimate the heat dissipated by the HOMs inside the CM, and thus the electrical power:

- $P_{elec\ Carnot} \propto 150 \times Q_{2K}$ ($Q_{2K}$: heat at 2K)

An order of magnitude of the maximum possible dissipation $Q_{2K}$ is 50-100 W, which could be impossible to withstand with the cryoplant.

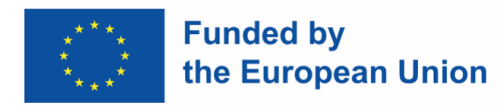

Funded by HORIZON-INFRA-2023-TECH-01 via grant agreement n°101131435

**b) With BLAs inside the CM at 50 K (BLA@50K)**

The power dissipated in BLAs by the HOMs is high (~50W between each cavity) but controlled. The BLAs are placed in between each cavity. Some absorber materials have a good efficiency when cooled at 50K (temperature loop provided by the cryoplant) but require R&D. When cooling the inter-cavity space at 50K, residual solid conduction heat is evacuated at 2K. The electrical power is estimated by:

- $P_{elec\ Carnot} \propto (150 \times Q_{2K} + 5 \times Q_{50K})$ ($Q_{50K}$ : heat at 50K)

**c) With BLAs inside the CM at 50K with thermalisation @5K (BLA@50K with therm. @5K)**

If an intermediate 5K interception loop is added to the previous scheme (BLA@50K), one can estimate the electrical power to be:

- $P_{elec\ Carnot} \propto (150 \times Q_{2K} + 60 \times Q_{5K} + 5 \times Q_{50K})$

BLA optimizations include the geometrical design (thermal design to reduce the $Q_{2K}$ with 50K-cooled BLAs) and R&D on microwave absorbing material cooled at 50K. **BLA will be placed in the CM. Depending on the final inter-cavity and the BLA mechanical implementation design, the 5K interception could be avoided** to circumvent the higher complexity and the transfer line heat losses brought by an additional 5K loop inside the CM.

## 6. ENERGY RECOVERY LINAC

Energy recovery linac (ERLs) are a promising technology and accelerator concepts to provide high current, high brightness, and high-quality beam by minimizing the accelerator footprint and the power consumption. One key aspect is that the ERL concept enables to recover the beam power of the accelerating beam and therefore to reduce the RF requirements compared to “classical” acceleration. To provide an order of magnitude of this power saving, one can consider the ERL of the PERLE [9] project as a reference example. PERLE aims at providing a CW 20 mA – 250 MeV (5 MW) electron beam at the Interaction point (IP) (cf. Figure 7). The ERL will need one cryomodule composed of 4 cavities, each cavity providing an accelerating voltage of 20.5 MV.

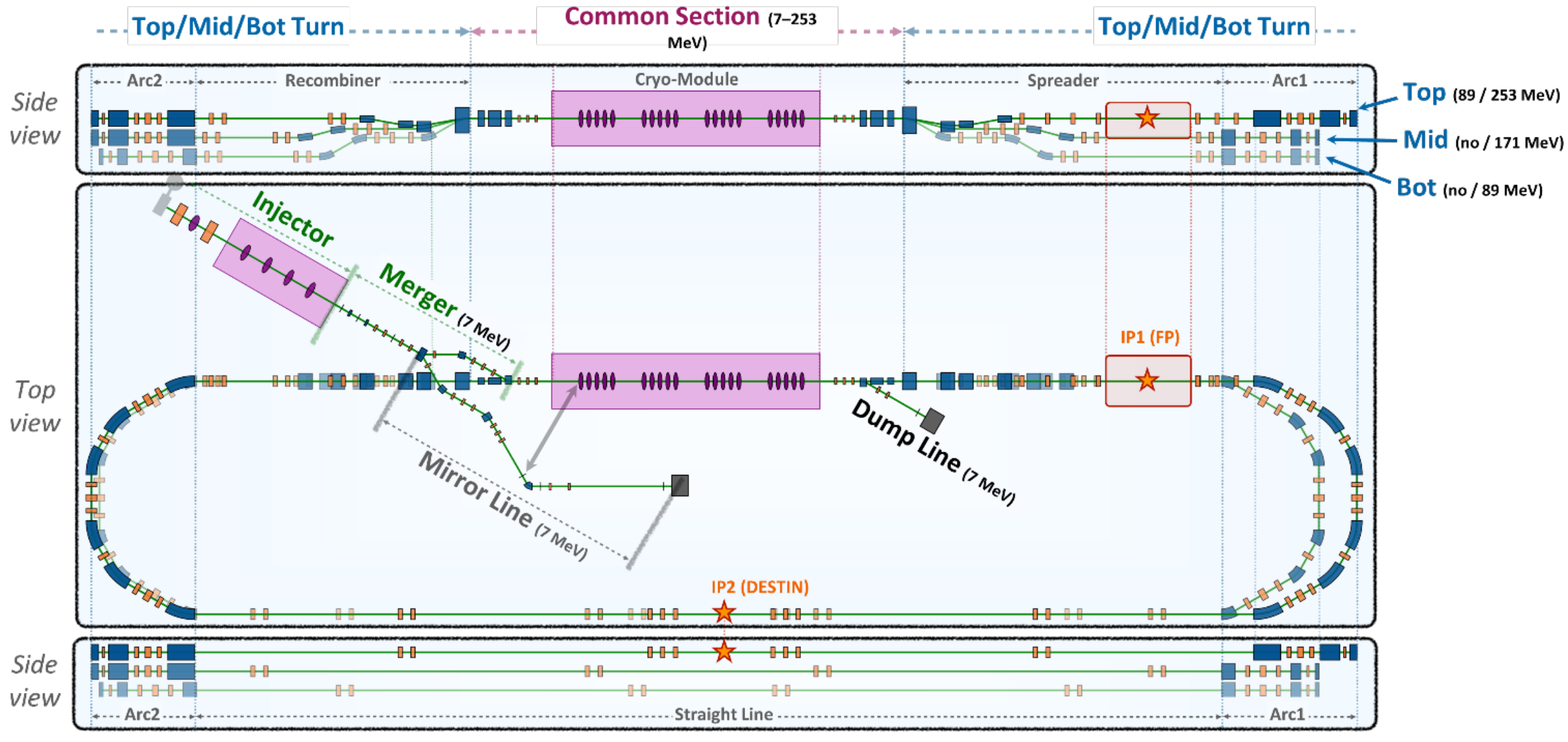

*Figure 7: Synoptic view of PERLE [10].*

In nominal ERL operation mode, some bunches are accelerated by the cavities and require power from the system (RF amplifier+ cavity]), while other bunches are decelerated and feed the system with their power. Consequently, one can show that in nominal ERL mode, the beam current seen by the system is 0 mA. Therefore, the $Q_{ext}$ value has a strong impact on the RF consumption and the cavity bandwidth. As shown by figures 8 and 9, a choice for $Q_{ext}$~$8.10^6$ would enable to provide a 100 Hz bandwidth for a reliable control of the field in the cavity with the LLRF. This means that each cavity will have to be operated with a minimum RF power of ~30 kW.

In the case of a linac providing the same beam as PERLE at the IP, it would require 3 cryomodules of 4 cavities and the minimum power to be provided to each cavity should be ~410 kW (see Figure 8). Consequently, one can evaluate, at first order, the gain on the total RF power to accelerate a 20-mA beam from 7 MeV to 250 MeV:

- ERL: Minimum RF power required ~4 x 30 kW = 120 kW
- Linac: Minimum RF power required ~ 4 x 3 x 410 kW = 4920 kW.

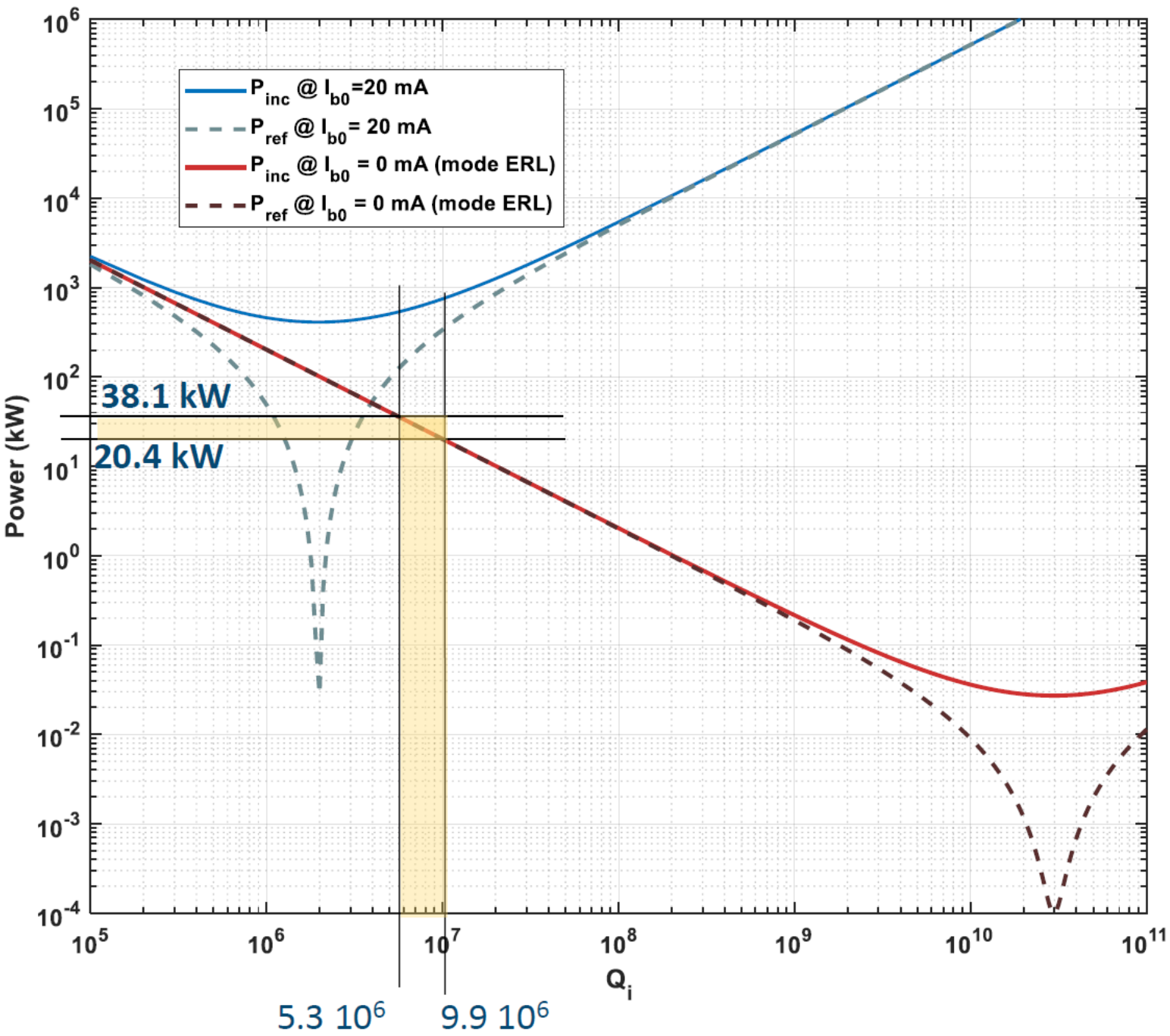

*Figure 8: Forward power required – and reflected power - as function of Qext for PERLE cavities. Two modes are presented: classical linac acceleration (current seen by the cavity: 20 mA) and ERL mode (current seen by the cavity: 0 mA) [11].*

Funded by the European Union

Funded by HORIZON-INFRA-2023-TECH-01 via grant agreement n°101131435

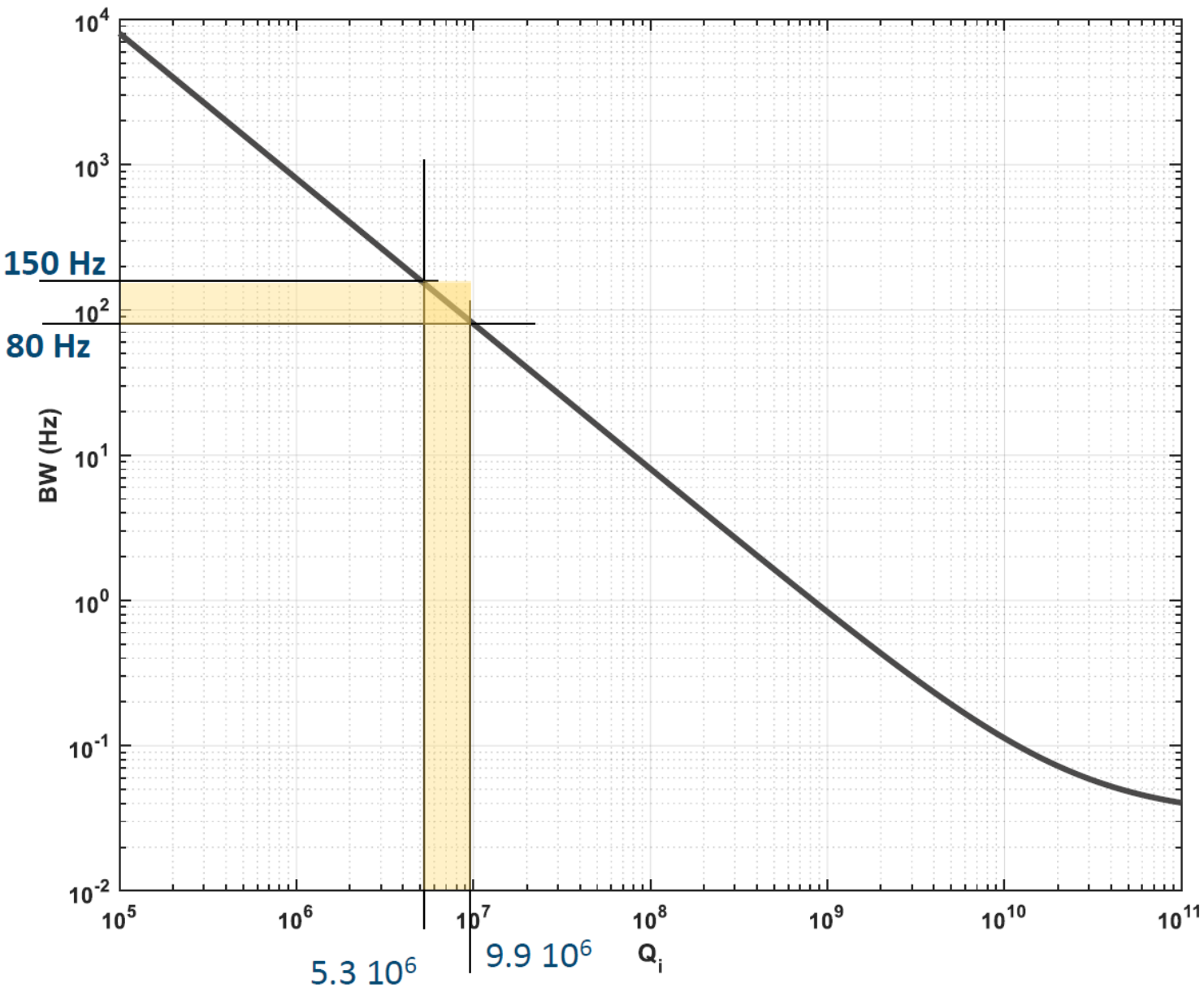

*Fig. 9: Coupling choice and cavity bandwidth.*

Nevertheless, for an ERL application, the cryomodule should enable to accelerate a high current beam with high efficiency and strong HOM damping capabilities. The cavities should operate with a high $Q_0$ (typically $3.10^{10}$) to minimize cryogenics dissipation. One should also take into account the complexity of an ERL that requires much more magnetic elements than in a linac.

**For the metric,** a consumption model of the PERLE ERL will be developed to evaluate the accelerator efficiency. The goal is to evaluate this beam power at the IP as a fraction of total electricity consumption. This efficiency will be calculated with the best accuracy considering all the elements of the machine: electron gun & associated systems (such as laser), RF amplifiers, the cryogenic plant, magnets, diagnostics and ancillary systems (vacuum, control, cooling, etc.). Impacts of errors due to perturbations (i.e. microphonics) or synchronisation (path length adjustment) will also be evaluated. The same exercise will be applied to a classical accelerator solution (such as a linac) and compared to the ERL efficiency.

In a second step, the impact of the iSAS technologies to improve the PERLE ERL efficiency could be evaluated:

- Ferro-Electric Fast Reactive Tuners: Could enable to operate with a higher $Q_{ext}$ and therefore to decrease the RF power consumption,
- Low Level Radio Frequency control system: could enable to decrease of LLRF electrical consumption,
- $Nb_3Sn$ on Cu films cavity: could enable to operate at 4 K instead of 2 K and therefore decrease the electrical consumption of the cryogenic plant,
- Couplers (FPC, HOM, BLA): could enable to minimise electrical power consumption of the cooling systems and cryogenic plant.

Funded by the European Union

Funded by HORIZON-INFRA-2023-TECH-01 via grant agreement n°101131435

## 7. SUMMARY OF METRIC

In summary, the following use cases will be evaluated at the end of iSAS:

**WP1 FE-FRT: measurements of RF power (TESLA cavity in CW mode at 16-20 MV/m) compared to a reference case (to be converted to wall plug power)**

- Reference test 1: conservative $Q_L = 1 \cdot 10^7$ with piezo system
- Reference test 2: high $Q_L = 5 \cdot 10^7$ with piezo system
- FE-FRT case test 3: low and high $Q_L$ with FE-FRT

**WP2 LLRF: calculations and measurements in different facilities/test conditions (CW, pulsed …)**

- Measure the accelerator AC consumption with and w/o the applied LLRF optimization (instantaneous AC consumption)
- Measure the peak and the integrated forward RF power with and w/o the LLRF optimization (RF power usage)
- Compute the AC-to-RF efficiency of the high-power source (i.e. SSA) with and w/o working point optimization of LLRF
- Provide examples of improved up-time or trip recovery time thanks to LLRF supervisory control and fault diagnostics

**WP3 $Nb_3Sn$ on Cu: comparison of $P_{acc}$ per cell at 1MV/m for $Nb_3Sn$ and for bulk Nb**

- Measurement of $Q_0$ *vs* $E_{acc}$ via an RF test ➔ deduce the surface resistance
- From the surface resistance of $Nb_3Sn$, estimate the operational (and construction) cost at 4K and compare to bulk Nb 2K same gradient
- Study cases
  - QPR data already available
  - 1.3 GHz cavity
  - new cavity optimized at the end of ISAS

**WP4/5 FPC: calculation of electrical power consumption with existing coupler design (800 MHz)**

- Comparison between cooling regime (turbulent/laminar) with existing design
- Electrical consumption considering the inlet line to couplers with feedback from ESS
- Comparison between a two-loop design (50K & 5K) and the existing one-loop design

**WP4 HOM: calculation of the electrical power consumption (800 MHz)**

- Comparison between 2 schemes: HOM@2K and HOM@5K

**WP4 BLA: calculation of the electrical power consumption (800 MHz)**

- Comparison between 3 schemes: no cold BLA (BLA@300K), BLA@50K and BLA@50K with thermalization @5K

**ERL: case of PERLE**

- Evaluation of the beam power at the IP as a fraction of total electricity consumption for PERLE and comparison with a classical accelerator

Funded by the European Union

Funded by HORIZON-INFRA-2023-TECH-01 via grant agreement n°101131435